\documentclass{article}
\pdfoutput=1

\input{defs.sty}

\renewcommand{\hypnamref}[2][empty]{\hyperref[#1]{#2}}
\renewcommand{\AllRefCited}[1]{}
\renewcommand{\seedefTotient}{}
\renewcommand{\includereduc}{\includegraphics[width=\textwidth]{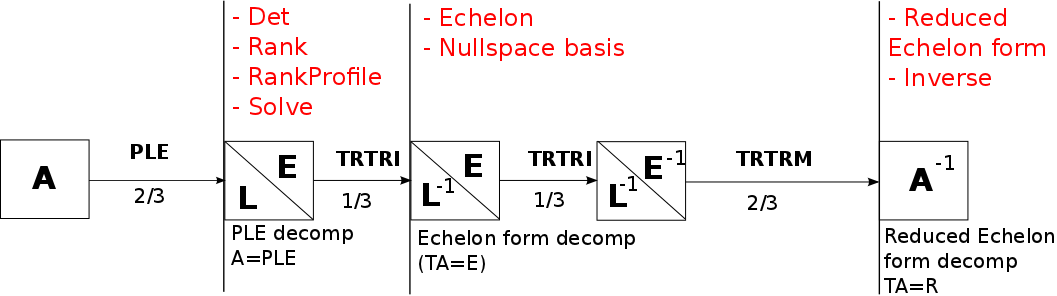}}

\usepackage{amssymb}
\usepackage{amsmath,amsfonts}

\newtheorem{theorem}{Theorem}
\newtheorem{definition}[theorem]{Definition}

\newtheorem{remark}[theorem]{Remark}

\usepackage{float}
\floatstyle{ruled}
\newfloat{algorithmfloat}{htbp}{loa}[section]
\floatname{algorithmfloat}{Algorithm}
 \newenvironment{algorithm}[1]{
   \begin{algorithmfloat}{\caption{#1}}
 }{\end{algorithmfloat}}

\newcommand{\F}{{\mathbb F}}
\newcommand{\Z}{{\mathbb Z}}

\let\oldsection=\section
\let\oldsubsection=\subsection

\renewcommand{\subsubsection}[1]{\oldsubsection{#1}}
\renewcommand{\subsection}[1]{\oldsection{#1}}
\renewcommand{\section}[1]{\title{#1}}

\newcommand{\sectionauthor}[2]{}
\newcommand\Referencescited{{\bf  References Cited:}\ }

\usepackage{graphicx,url,color,svgcolor}

\makeatletter

\usepackage{hyperref}
\hypersetup{
pdftitle={\@title},
pdfauthor={\@author},
breaklinks=true, %permet le retour à la ligne dans les liens trop longs
colorlinks=true,
 linkcolor=blue,
 citecolor=darkred,
 urlcolor=darkblue,
}

\newenvironment{sectionauthors}{}{
\author{Jean-Guillaume Dumas\footnote{Universit\'e de Grenoble;
    Laboratoire Jean Kuntzmann, (umr CNRS 5224, Grenoble INP, INRIA,
    UJF, UPMF);
    \href{mailto:Jean-Guillaume.Dumas@imag.fr}{Jean-Guillaume.Dumas@imag.fr};
    51, rue des Math\'ematiques, 
    BP 53X, F-38041 Grenoble, France.
  } \and Cl{\'e}ment Pernet\footnote{INRIA, Universit\'e de Grenoble;
    Laboratoire LIG (umr CNRS 5217, Grenoble INP, INRIA, UJF, UPMF);
    \href{mailto:Clement.Pernet@imag.fr}{Clement.Pernet@imag.fr};
    ENSIMAG Antenne de Montbonnot, 
    51, avenue Jean Kuntzmann, 
    F-38330 Montbonnot Saint-Martin, France.
  }
}
\maketitle
\setcounter{tocdepth}{5}
\@starttoc{toc}
\newpage
}
\renewenvironment{table}[1][empty]
               {\@float{table}[#1]\begin{center}}
               {\end{center}\end@float}
\makeatother

\usepackage[english]{babel}
\usepackage{babelbib}

\begin{document} 
\section{\titlename}\label{sec134}

\begin{sectionauthors}
\sectionauthor{Jean-Guillaume Dumas}{Universit{\'e} de Grenoble}
\sectionauthor{Cl{\'e}ment Pernet}{Universit{\'e} de Grenoble}
\end{sectionauthors}

% Most implementations are in LinBox.
We present here algorithms for efficient computation of
linear algebra problems over finite fields. 
Implementations\footnote{\url{http://magma.maths.usyd.edu.au},
  \url{http://www.maplesoft.com}, \url{http://sagemath.org},
  \url{http://www.shoup.net/ntl},
  \url{http://www.flintlib.org}, 
  \url{http://www.cs.uwaterloo.ca/~astorjoh/iml.html},
  \url{http://m4ri.sagemath.org},
  \url{http://linalg.org}}
of the proposed algorithms are available through the {\sc Magma}, {\sc
  Maple} (within the \verb!LinearAlgebra[Modular]! subpackage) and
{\sc Sage} systems; some parts can also be found within the C/C++ libraries
NTL,
FLINT, IML, M4RI and the special purpose {\sc LinBox} template library
for exact, high-performance linear algebra computation with dense,
sparse, and structured matrices over the integers and over finite
fields \cite{DumGauGieGioHovKalSauTurVil02}.  

\subsection{Dense matrix multiplication}\label{ssec:blas}

\begin{definition} \label{def:mm}
For $A \in \F_q^{m\times k}$ and $B \in \F_q^{k\times n}$ with elements $A_{i,j}$ and $B_{i,j}$,
the matrix $C=A\times B$ has $C_{i,j} = \sum_{l=1}^{k} A_{i,l}
B_{l,j}$. We denote by $\MatrixMul(m,k,n)$ a time complexity bound on the
number of field operations necessary to compute $C$.
\end{definition}
Classical triple loop implementation of matrix multiplication makes
$\MatrixMul(m,k,n)\leq2mkn$. The best published estimates to date
gives $\MatrixMul(n,n,n) \leq \BigO{n^\omega}$ with $\omega \approx
2.3755$ \cite{MR1056627}, though improvements to $2.3737$ and $2.3727$
are now claimed \cite{Sto10,Vas11}. For very rectangular matrices one
also have astonishing results like $\MatrixMul(n,n,n^\alpha) \leq
\BigO(n^{2+\epsilon})$ for a constant $\alpha > 0.294$ and any
$\epsilon>0$ \cite{MR1449760}.
Nowadays practical implementations mostly use Strassen-Winograd's
algorithm, see section \ref{ssec:wino}, with an intermediate complexity and
$\omega \approx 2.8074$.
%$\omega$

\subsubsection{Tiny finite fields}\label{sssec:tiny}
The practical efficiency of matrix multiplication depends highly on
the representation of field elements. 
We thus present three kinds of compact representations for elements of a
finite field with
very small cardinality: bitpacking (for $\F_2$), bit-slicing (for say
$\F_3, \F_5, \F_7, \F_{2^3}$, or $\F_{3^2}$) and Kronecker
substitution\index{Kronecker!substitution}. These representations are
designed to allow
efficient linear algebra operations, including matrix multiplication.
\index{Four Russians (Method!of)}
\index{bit-packing}

% Definition ou Prop?

\begin{algorithm}{[Greasing]}\label{alg:Greasing}\index{Greasing}
Over $\F_2$, the method of \textit{the four Russians}~\cite{MR0269546}, also called
\textit{Greasing} can be used as follows:
\begin{itemize}
\item A 64 bit machine word can be used to represent a row vector of dimension~64.
\item Matrix multiplication of a $m\times k$ matrix  $A$ by a $k\times n$
matrix $B$ can be done by first storing all $2^k$ $k$-dimensional linear
combinations of rows of $B$ in a table. Then the i-th row of the
product is copied from the row of the table indexed by the i-th row of $A$.
\item By ordering indices of the table according to a binary Gray Code, each
row of the table can be deduced from the previous one, using only one row
addition. This brings the bit operation count to build the table from $k2^kn$ to $2^kn$.
\item Choosing $k=\log_2n$ in the above method implies $\MatrixMul(n)=\BigO{n^3/\log n}$
over~$\F_2$.
\end{itemize}
\end{algorithm}

\index{bit-slicing}
\begin{definition}\cite{BooBra09}
Bitslicing consists in representing an $n$-dimensional vector of $k$-bit sized
coefficients using $k$ binary vectors of dimension $n$. In particular, one can
use boolean word instruction to perform arithmetic on 64 dimensional vectors.
\begin{itemize}
\item Over $\F_3$, the binary representation $0 \equiv [0,0], 1\equiv
[1,0], -1 \equiv [11]$ allows to add and subtract two elements in 6 boolean operations:
$$\begin{array}{ll}
\text{Add}([x_0,x_1],[y_0,y_1]) : & s \leftarrow x_0\oplus y_1 , t \leftarrow x_1 \oplus y_0\\
      & \text{Return} (s\wedge t, (s\oplus x_1) \vee (t\oplus y_1))\\
\text{Sub}([x_0,x_1],[y_0,y_1]) : & t\leftarrow x_0\oplus y_0 \\
                                 & \text{Return}(t\vee (x_1\oplus y_1), (t\oplus
                                 y_1)\wedge(y_0\oplus x_1))
\end{array}
$$
\item Over $\F_5$ (resp. $\F_7$), a redundant representation $x=x_0+2x_1+4x_2 \equiv
[x_0,x_1,_2]$ allows to add two elements in 20 (resp. 17) boolean operations,
negate in 3 (resp. 6) boolean operations and double in 0 (resp. 5) boolean operations.
\end{itemize}
\end{definition}

\begin{table}[htbp]
\begin{tabular}{l|ccc}
         & $\F_3$ & $\F_5$ & $F_7$\\
Addition & 6 & 20 & 17\\
Negation & 1 & 5 &  3\\
Double   &   & 5 &  0\\
\end{tabular}
\caption{boolean operation counts for basic arithmetic using bit slicing}

\end{table}
%%% A reprendre en étant plus explicite sur la différence entre
%%% bit-packing/slicing et kronecker substitution

% CP: me suis permis de modifier ta definition pour rendre le distingo plus
% clair avec bitslicing
\index{bit-packing}
\begin{definition}
Bitpacking consists in representing a vector of field elements as an integer
fitting in a single machine word using a $2^k$-adic
representation: $$(x_0,\dots,x_{n-1})\in \F_q^n \equiv X=x_0+2^kx_1+\dots
+(2^k)^{n-1}x_{n-1} \in \Z_{2^{64}}$$ 
%%  tiny finite fields, it is possible to pack several field elements
%% into a single machine word. 
Elements of extension fields are viewed as
polynomials and stored as the evaluation of this polynomial at the
characteristic of the field. The latter evaluation is called {\em
  Kronecker substitution}\index{Kronecker!substitution}.
\end{definition}
We first need a way to simultaneously reduce coefficients modulo the
characteristic, see \cite{MR2500374}.
\begin{algorithm}{[REDQ: Q-adic REDuction]}\label{alg:REDQ}\index{REDQ}
\begin{algorithmic}[1]
\REQUIRE Three integers $p$, $q$ and $\tilde{r}  = \sum_{i=0}^d \widetilde{\mu_i} q^i \in \Z$.
\ENSURE $\rho \in \Z$, with $\rho = \sum_{i=0}^d \mu_i q^i$
where $\mu_i = \widetilde{\mu_i} \bmod p$.
\UNDERL{REDQ COMPRESSION\index{REDQ!Compression}}
\STATE $s = \left\lfloor \frac{\tilde{r}}{p} \right\rfloor$;
\FOR{$i=0$ to $d$}
\STATE $u_i = \left\lfloor \frac{\tilde{r}}{q^i} \right\rfloor -  p \left\lfloor \frac{s}{q^i} \right\rfloor$;
\ENDFOR
\UNDERL{REDQ CORRECTION\index{REDQ!Correction}} \hfill\COMMENT{only when $p\nmid q$, otherwise $\mu_i=u_i$ is correct}
\STATE $\mu_{d}=u_{d}$;
\FOR{$i=0$ to $d-1$}
\STATE $\mu_i = u_i-qu_{i+1} \bmod p$;
\ENDFOR
\\[5pt]\STATE Return $\rho = \sum_{i=0}^d \mu_i q^i$;
\end{algorithmic}
\end{algorithm}

Once we can pack and simultaneously reduce coefficients of finite
field in a single machine word, the obtained parallelism can be used
for matrix multiplication. Depending on the respective sizes of the
matrix in the multiplication one can pack only the left operand or
only the right one or both \cite{DumFouSal11}. We give here only a generic
algorithm for packed matrices, which use multiplication of a right
packed matrix by a non packed left matrix.
\begin{algorithm}{[Right packed matrix
  multiplication\index{packed!matrix multiplication}]}\label{alg:rightcomp}
\begin{algorithmic}[1]
\REQUIRE A prime $p$ and $A_c \in \F_p^{m\times k}$ and $B_c \in \F_p^{k\times n}$,
stored with several field elements per machine word.
\ENSURE $C_c=A_c\times B_c\in \F_p^{m\times n}$
\STATE $A=\operatorname{Uncompress}(A_c)$; \hfill\COMMENT{extract the
  coefficients}
\STATE $C_c=A\times B_c$; \hfill\COMMENT{Using e.g., algorithm
  \ref{alg:fgemm}}
\STATE Return $\operatorname{REDQ}(C_c)$;
\end{algorithmic}
\end{algorithm}

Then, over extensions, fast floating point operations can be used on
the Kronecker substitution\index{Kronecker!substitution} of the
elements. 
Indeed, it is very often desirable to use floating point
arithmetic, {\em exactly}.
For instance floating point
routines can more easily use large hardware registers, they can more easily
optimize the memory hierarchy usage \cite{MR2501869,WhaPetDon01} and
portable implementations are more widely available. 
We present next the dot product and the matrix
multiplication is then straightforward
\cite{MR2035233,MR2500374,DumFouSal11}. 
\begin{algorithm}{[Compressed Dot product over extension fields]\label{alg:FGDP}}
\begin{algorithmic}[1]
\REQUIRE A field $\F_{p^k}$ with elements represented as exponents of
a generator of the field;
\REQUIRE two vectors $v_1$ and $v_2$ of elements of $\F_{p^k}$;
\REQUIRE a sufficiently large integer $q$.
\ENSURE $R \in \F_{p^k}$, with $R = v_1^T\cdot v_2$.
\newline{\COMMENT{Tabulated conversion: uses tables from exponent to floating point evaluation}}
\STATE Set $\widetilde{v_1}$ and $\widetilde{v_2}$ to the floating
point Kronecker substitution\index{Kronecker!substitution} of the elements of $v_1$ and $v_2$.
\STATE Compute $\tilde{r} = \widetilde{v_1}^T \cdot\widetilde{v_2}$;
\hfill\COMMENT{The floating point computation}
\STATE $r = REDQ\_COMPRESSION(\tilde{r},p,q)$;\index{REDQ!Compression}
\hfill\COMMENT{Computing a radix
    decomposition}
\newline{\COMMENT{Variant of REDQ\_CORRECTION\index{REDQ!Correction}:
    $\mu_i = \widetilde{\mu_i} \bmod p$ for
$\tilde{r} = \sum_{i=0}^{2k-2} \widetilde{\mu_i} q^i$}}
\STATE Set $L = representation( \sum_{i=0}^{k-2} \mu_i X^i )$;
\STATE Set $H = representation( X^{k-1} \times \sum_{i=k-1}^{2k-2} \mu_i X^{i-k+1} )$;
\STATE Return $R = H + L \in \F_{p^k}$;
\hfill\COMMENT{Reduction in the field}
\end{algorithmic}
\end{algorithm}

\subsubsection{Word size prime fields}
\label{sec:fflas}
\index{BLAS}\index{FFLAS}
Over word-size prime fields one can also use the reduction to floating
point routines of algorithm \ref{alg:FGDP}.
The main point is to be able to perform efficiently the
matrix multiplication of blocks of the initial matrices without
modular reduction. Thus delaying the reduction as much as possible,
depending on the algorithm and internal representations, in
order to amortize its cost. We present next such a delaying with the
classical matrix multiplication algorithm and a centered
representation \cite{MR2738206}.

\begin{algorithm}{[{\texttt fgemm}: Finite Field GEneric Matrix Multiplication]\label{alg:fgemm}}
\begin{algorithmic}[1]
\REQUIRE An odd prime $p$ of size smaller than the floating point mantissa $\beta$
and $F_p$ elements stored by values between
$\frac{1-p}{2}$ and 
$\frac{p-1}{2}$
\REQUIRE $A \in \F_p^{m\times k}$ and $B \in \F_p^{k\times n}$
\ENSURE $C=A\times B\in \F_p^{m\times n}$
\IF {$n(p-1)^2<2^{\beta+1}$}
\STATE Convert $A$ and $B$ to floating point matrices $A_f$ and $B_f$;
\STATE Use floating point routines to compute $C_f=A_f \times B_f$;
\STATE $C = C_f \mod p$;
\ELSE
\STATE Cut $A$ and $B$ into smaller blocks;
\STATE Call the algorithm recursively for the block multiplications;
\STATE Perform the block additions modulo $p$;
\ENDIF
\end{algorithmic}
\end{algorithm}

\subsubsection{Large finite fields}\label{sssec:large}
If the field is too large for the strategy \ref{alg:fgemm} over
machine words, then two main approaches would have to be considered:
\begin{itemize}
\item Use extended arithmetic, either arbitrary of fixed
  precision, if the characteristic is large, and a polynomial
  representation for extension fields. 
  The difficulty here is to preserve an optimized memory
  management and to have an almost linear time extended precision polynomial
  arithmetic.
\item Use a residue number system and an evaluation/interpolation
  scheme: one can use algorithm \ref{alg:fgemm} for each prime in the
  RNS and each evaluation point. For $\F_{p^k}$, the number of needed
  primes is roughly $2\log_{2^\beta}(p)$ and the number of evaluations
  points is $2k-1$.
\end{itemize}

\subsubsection{Large matrices: subcubic time complexity}\label{ssec:wino}

With matrices of large dimension, sub-cubic time complexity algorithms, such as
Strassen-Winograd's~\cite{MR0297115} can be used to decrease the number of
operations. Algorithm~\ref{alg:strwin} describes how to compute one recursive
level of the algorithm, using seven recursive calls and 15 block additions.

\begin{algorithm}{[Strassen-Winograd]}
\label{alg:strwin}
%\begin{algorithmic}
%\end{algorithmic}
%\vspace{-2pt}%{
%\newline
%\begin{minipage}{.3\columnwidth}
$$A=\begin{bmatrix}
  A_{11} & A_{12}\\
  A_{21} & A_{22}\\
\end{bmatrix};
 B=
\begin{bmatrix}
  B_{11} & B_{12}\\
  B_{21} & B_{22}\\
\end{bmatrix};
C=\begin{bmatrix}
  C_{11} & C_{12}\\
  C_{21} & C_{22}\\
\end{bmatrix};$$
%\end{minipage}
%\begin{minipage}{.7\columnwidth}

$$
\begin{array}{lllll}
 S_1  \leftarrow A_{21} + A_{22};  &  T_1 \leftarrow B_{12} - B_{11}; & P_1 \leftarrow A_{11} \times B_{11}; & P_2 \leftarrow A_{12} \times B_{21};\\
 S_2  \leftarrow S_1 - A_{11};   &  T_2 \leftarrow B_{22} - T_1;   & P_3  \leftarrow S_4 \times B_{22};  &P_4  \leftarrow A_{22} \times T_4;\\
 S_3  \leftarrow A_{11} - A_{21};  &  T_3 \leftarrow B_{22} - B_{12}; & P_5 \leftarrow S_1 \times T_1; & P_6\leftarrow S_2 \times T_2; \\
 S_4  \leftarrow A_{12} - S_2;	& T_4 \leftarrow T_2 - B_{21};   &  P_7 \leftarrow S_3 \times T_3; \\
\end{array}$$
%\end{minipage}
%}
%\vspace{-10pt}
% $\begin{array}{lll}
%   S_1  \leftarrow A_{21} + A_{22}	& \hspace{2em}  & T_1 \leftarrow B_{12} - B_{11} \\
%   S_2  \leftarrow S_1 - A_{11}		&				& T_2 \leftarrow B_{22} - T_1	 \\
%   S_3  \leftarrow A_{11} - A_{21}	&				& T_3 \leftarrow B_{22} - B_{12} \\
%   S_4  \leftarrow A_{12} - S_2		&				& T_4 \leftarrow T_2 - B_{21}	 \\
% \end{array}$\\
% \[
% \begin{array}{llll}
%   T_1 \gets B_{12} - B_{11}  & T_2  \leftarrow B_{22} - T_1	& T_3 \leftarrow B_{22} - B_{12}	& T_4 \leftarrow T_2 - B_{21}	 \\
%   S_1 \leftarrow A_{21} + A_{22}  & S_2  \leftarrow S_1 - A_{11}	& S_3 \leftarrow A_{11} - A_{21}	& S_4  \leftarrow A_{12} - S_2  
% \end{array}
% \]
%
%\item 7 recursive multiplications:
%\vspace{-2pt}%{ 
%}
%\vspace{-10pt}
% \[
% $\begin{array}{lll}
%   P_1  \leftarrow A_{11} \times B_{11}  &  \hspace{2em} & P_5 \leftarrow S_1 \times T_1 \\
%   P_2  \leftarrow A_{12} \times B_{21}  &				& P_6\leftarrow S_2 \times T_2	\\
%   P_3  \leftarrow S_4 \times B_{22}     &				& P_7 \leftarrow S_3 \times T_3 \\
%   P_4  \leftarrow A_{22} \times T_4														\\
% \end{array}$
% \]
%
%\item 7 final additions:
%\vspace{-2pt}%{
$$\begin{array}{llll}
C_{11}  \leftarrow P_1 + P_2; & U_2  \leftarrow P_1 + P_6; & U_3  \leftarrow U_2 +
P_7; & U_4  \leftarrow U_2 + P_5;\\
C_{12} \leftarrow U_4 + P_3; & C_{21} \leftarrow U_3 - P_4;& C_{22} \leftarrow U_3 + P_5;  \\
\end{array}$$
%}
%\vspace{-10pt}
% \[
% \begin{array}{lll}
%   U_1  \leftarrow P_1 + P_2 & \hspace{2em}  & U_5 \leftarrow U_4 + P_3  \\
%   U_2  \leftarrow P_1 + P_6 &				& U_6 \leftarrow U_3 - P_4	\\
%   U_3  \leftarrow U_2 + P_7 &				& U_7 \leftarrow U_3 + P_5  \\
%   U_4  \leftarrow U_2 + P_5											    \\
% \end{array}
% \]
%
%\item The result is the matrix:
%{ $ C = \left[ \begin{array}{ll} U_1 & U_5 \\ U_6 & U_7 \end{array} \right]$}.
%\end{itemize}
%\end{inparaenum}
\end{algorithm}

In practice, one uses a threshold in the matrix dimension to switch to a
base case algorithm, that can be any of the one previously described.
% For instance, over $\F_{1000\,003}$ using double precision floating point
% representation, and ATLAS BLAS on a 64 bit Intel-i7, the threshold is at
% $n=768$. With $n=5000$, the improvement is of $23.8\%$ compared to classical
% matrix multiplication.
Following section~\ref{sec:fflas}, one can again delay the
modular reductions, 
but the intermediate computations of Strassen-Winograd's algorithm impose a
tighter bound:

\begin{theorem}\cite{MR2738206}
Let $A \in \Z^{m \times k}$, $B \in \Z^{k \times n}$  $C \in \Z^{m
  \times n}$ and $\beta \in \Z$ with

%% $ m_A \leq a_{i,j} \leq M_A $, $m_B \leq b_{i,j} \leq M_B$ and  $m_C \leq
%% c_{i,j} \leq M_C$. Moreover, suppose that $0\leq -m_A \leq M_A$,  $0\leq -m_B \leq
%% M_B$, $0\leq -m_C \leq M_C$, $M_C \leq M_B$ and $|\beta| \leq M_A, M_B$.
$a_{i,j},b_{i,j},c_{i,j},\beta \in \{0\dots p-1\}$. 
%we have
% $ 0 \leq a_{i,j} < p $, $0 \leq b_{i,j} < p$, $0 \leq c_{i,j} < p$
%and $0\leq \beta < p$.
Then 
every intermediate value $z$ involved in the computation of  $A \times B +
\beta C$ with $l$ ($l\geq 1$) recursive levels of algorithm~\ref{alg:strwin} satisfy: 
\[ \left|z \right| \leq \left(\frac{1+3^l}{2}\right)^2 \left\lfloor{
    \frac{k}{2^l}}\right\rfloor (p-1)^2 \]

Moreover, this bound is tight.
\end{theorem}
For instance, on a single Xeon 2.8GHz core with gcc-4.6.3, Strassen-Winograd's
variant implemented with LinBox-1.2.1 and GotoBLAS2-1.13 
%can be 45\% faster for the multiplication of $20\,000\times 20\,000$
%matrices over $\F_{2^{19}-1}$, in less than $13$ minutes.
can be 37\% faster for the multiplication of $10\,000\times 10\,000$
matrices over $\F_{2^{19}-1}$, in less than $1'49"$.
\subsection{Dense Gaussian elimination and echelon
  forms}\label{ssec:echelon}
In this section, we present algorithms computing the determinant and
inverse of square matrices; the rank, rank profile, nullspace, and
system solving for arbitrary shape and rank matrices. All these
problems are solved a la Gaussian elimination, but recursively in
order to effectively incorporate matrix multiplication. The latter is
denoted generically \texttt{gemm} and, depending on the underlying
field, can be implemented using any of the techniques of sections
\ref{sssec:tiny}, \ref{sec:fflas} or \ref{sssec:large}.

 A special care
is given to the asymptotic time complexities: the exponent is reduced to that of matrix
multiplication using block recursive algorithms, and the constants are also
carefully compared. Meanwhile, this approach is also effective for
implementations: grouping arithmetic operations into matrix-matrix products allow
to better optimize cache accesses.

\subsubsection{Building blocks}
Algorithms~\ref{alg:trsmrec}, \ref{alg:trmm}, \ref{alg:trtri} and~\ref{alg:trtrm} show how to reduce the
computation of triangular matrix systems, triangular matrix multiplications, and
triangular matrix inversions to matrix-matrix multiplication. Note that they do
not require any temporary storage other than the input and output arguments.

%\cite{MR2738206}
\begin{algorithm}{[\texttt{trsm}: Triangular System Solve with Matrix
    right hand side]\label{alg:trsmrec}}
\begin{algorithmic}[1]
\begin{minipage}{0.58\columnwidth}
\REQUIRE{$A \in \F_q^{m \times m}$ non-singular upper triangular, $B \in \F_q^{m \times n}$}
\ENSURE{$X \in \F_q^{m \times n}$ s.t. $AX= B$}
\IFTHEN{ m=1 }{{\algorithmicreturn} $X= A_{1,1}^{-1} \times B$}
%\tcc{ (découpage des matrices en blocs de tailles $\left\lfloor \frac{m}{2}
% \right\rfloor$ et $\lCeil \frac{m}{2} \rCeil$) }\;
  \STATE  $X_2=${\tt trsm($A_3,B_2$)};
  \STATE $B_1= B_1 - A_2X_2$; \COMMENT{using
    \hypnamref[alg:fgemm]{gemm}, e.g., via alg. \ref{alg:fgemm}}
  \STATE $X_1=${\tt trsm($A_1,B_1$)}; 
\RETURN $X=
\begin{bmatrix}
X_1 \\ X_2\end{bmatrix};
$
\end{minipage}\hfill
\begin{minipage}{0.36\columnwidth}
Using the conformal block \mbox{decomposition}:\\ 
$
  \begin{bmatrix}
  A_1&A_2\\&A_3
  \end{bmatrix}
  \begin{bmatrix}
  X_1\\X_2
  \end{bmatrix}
  =
  \begin{bmatrix}
  B_1\\B_1
  \end{bmatrix}
  $
\end{minipage}
\end{algorithmic}
\end{algorithm}
\index{TRSM}
%% \begin{theorem} \label{thm:TRSM}
%% Algorithm TRSM is correct and the leading term of its arithmetic
%% complexity is
%% $TRSM(m,n) = \frac{1}{2^{\omega-1}-2}\lceil\frac{n}{m}\rceil  MM(m).$
%% This complexity is
%% $m^2n$
%% using classic matrix  multiplication.
%% \end{theorem} 

\begin{algorithm}{[\texttt{trmm}: Triangular Matrix Multiplication]\label{alg:trmm}}
\begin{algorithmic}[1]
\begin{minipage}{0.58\columnwidth}
\REQUIRE{ $A \in \F_q^{m \times m}$ upper triangular, $B \in \F_q^{m \times n}$}
\ENSURE{$C \in \F_q^{m \times n}$ s.t. $AB= C$}
\IFTHEN{ m=1 }{{\algorithmicreturn} $C= A_{1,1} \times B$}
  \STATE $C_1=${\tt trmm($A_1,B_1$)}; 
  \STATE $C_1= C_1 + A_2B_2$; \COMMENT{using \hypnamref[alg:fgemm]{gemm}}
  \STATE $C_2=${\tt trmm($A_3,B_2$)}; 
\RETURN $C=
\begin{bmatrix}
C_1 \\ C_2\end{bmatrix};
$
\end{minipage}\hfill
\begin{minipage}{0.36\columnwidth}
%\tcc{ (découpage des matrices en blocs de tailles $\left\lfloor \frac{m}{2}
% \right\rfloor$ et $\lCeil \frac{m}{2} \rCeil$) }\;
Using the conformal block \mbox{decomposition}:\\ 
$
  \begin{bmatrix}
  A_1&A_2\\&A_3
  \end{bmatrix}
  \begin{bmatrix}
  B_1\\B_2
  \end{bmatrix}
  =
  \begin{bmatrix}
  C_1\\C_2
  \end{bmatrix}
  $
\end{minipage}
\end{algorithmic}
\end{algorithm}

\begin{algorithm}{[\texttt{trtri}: Triangular Matrix Inversion]\label{alg:trtri}}
\begin{algorithmic}[1]
\begin{minipage}{0.58\columnwidth}
\REQUIRE{ $A \in \F_q^{n \times n}$ upper triangular and non-singular}
\ENSURE{$ C=A^{-1}$}
\IFTHEN{ m=1 }{{\algorithmicreturn} $ C= A_{1,1}^{-1}$}
  \STATE $C_1=A_1^{-1}$; \COMMENT{using \hypnamref[alg:trtri]{trtri} recursively}
  \STATE $C_3=A_3^{-1}$; \COMMENT{using \hypnamref[alg:trtri]{trtri} recursively}
  \STATE $C_2=A_2C_3$; \COMMENT{using  \hypnamref[alg:trmm]{trmm} }
  \STATE $C_2=-C_1C_2$; \COMMENT{using \hypnamref[alg:trmm]{trmm} }
\RETURN $C=
\begin{bmatrix}
C_1 & C_2\\&C_3\end{bmatrix};
$
\end{minipage}\hfill
\begin{minipage}{0.36\columnwidth}
%\tcc{ (découpage des matrices en blocs de tailles $\left\lfloor \frac{m}{2}
% \right\rfloor$ et $\lCeil \frac{m}{2} \rCeil$) }\;
Using the conformal block \mbox{decomposition}:\\
$
  \begin{bmatrix}
  A_1&A_2\\&A_3
  \end{bmatrix},
  \begin{bmatrix}
  C_1&C_2\\&C_3
  \end{bmatrix}
  $
\end{minipage}
\end{algorithmic}
\end{algorithm}

\begin{algorithm}{[\texttt{trtrm}: Upper-Lower Triangular Matrix Multiplication]\label{alg:trtrm}}
\begin{algorithmic}[1]
\begin{minipage}{0.58\columnwidth}
\REQUIRE{ $L \in \F_q^{n \times n}$ lower triangular}
\REQUIRE{ $U \in \F_q^{n \times n}$ upper triangular}
\ENSURE{$ A=UL$}
\IFTHEN{ m=1 }{{\algorithmicreturn} $ A= U_{1,1}L_{1,1}$}
  \STATE $A_1=U_1L_1$; \COMMENT{using \hypnamref[alg:trtrm]{trtrm} recursively}
  \STATE $A_1=A_1+U_2L_2$; \COMMENT{using \hypnamref[alg:fgemm]{gemm}}
  \STATE $A_2=U_2L_3$; \COMMENT{using \hypnamref[alg:trmm]{trmm} }
  \STATE $A_3=U_3L_2$; \COMMENT{using \hypnamref[alg:trmm]{trmm} }
  \STATE $A_4=U_3L_3$; \COMMENT{using \hypnamref[alg:trtrm]{trtrm} recursively}
\RETURN $A=
\begin{bmatrix}
A_1 & A_2\\A_3&A_4\end{bmatrix};
$
\end{minipage}\hfill
\begin{minipage}{0.36\columnwidth}
%\tcc{ (découpage des matrices en blocs de tailles $\left\lfloor \frac{m}{2}
% \right\rfloor$ et $\lCeil \frac{m}{2} \rCeil$) }\;
Using the conformal block \mbox{decomposition}:\\ 
$\begin{bmatrix}
  L_1\\L_2&L_3
  \end{bmatrix},\begin{bmatrix}
  U_1&U_2\\&U_3
  \end{bmatrix},\begin{bmatrix}
  A_{1}&A_{2}\\A_{3}&A_{4}
  \end{bmatrix}$
\end{minipage}
\end{algorithmic}
\end{algorithm}

\subsubsection{PLE decomposition}\index{PLE decomposition}

Dense Gaussian elimination over finite fields can be reduced to matrix
multiplication, using the usual techniques for the LU decomposition of
numerical linear algebra~\cite{MR0331751}. 
However, in applications over a finite field, the input matrix often
has non-generic rank profile and special care needs to be taken about
linear dependencies and rank deficiencies.
The PLE decomposition is thus a generalization of the PLU
decomposition for matrices with any rank\index{Rank} profile. 
%, that occur commonly in applications
%over finite fields.
\begin{definition}
A matrix is in row-echelon form if all its zero rows occupy the last row
positions and the leading coefficient of any non-zero row except the first one is strictly to the right of the
leading coefficient of the previous row.
Moreover, it is said to be in reduced row-echelon form, if all coefficients
above a leading coefficient are zeros.
\end{definition}

\begin{definition}
For any matrix $A\in F_q^{m\times n}$ of rank\index{Rank} $r$, there is a PLE decomposition $A=PLE$
where $P$ is a permutation matrix, $L$ is a $m\times r$ lower triangular matrix
and $E$ is a $r\times n$ matrix in row-echelon form, with unit leading coefficients.
\end{definition}

Algorithm~\ref{alg:pledec} shows how to compute such a decomposition by a block
recursive algorithm, thus reducing the complexity to that of matrix
multiplication.

\begin{algorithm}{[PLE decomposition]\label{alg:pledec}}
\begin{algorithmic}[1]
\REQUIRE{ $A \in \F_q^{m \times n}$}
\ENSURE{$(P,L,E)$ a PLE decomposition of $A$}
\IF{$n=1$}
  \IFTHEN{$A=0_{m\times 1}$}{{\algorithmicreturn} $(I_m,I_0, A)$;}
     \STATE Let $j$ be the column index of the first non zero entry of $A$
     and $P=T_{1,j}$ the transposition between indices $1$ and $j$;
     \RETURN $(P,PA, [1])$;
\ELSE 
\begin{minipage}{\columnwidth}
\begin{minipage}{0.58\columnwidth}
   \STATE $(P_1,L_1,E_1) = \texttt{PLE}(A_1)$; \COMMENT{recursively}
   \STATE $A_2 = P_1A_2$;
   \STATE $A_3 = L_{1,1}^{-1}A_3$; \COMMENT{using \hypnamref[alg:trsmrec]{trsm}}
   \STATE $A_4 = A_4 - L_{1,2}A_3$; \COMMENT{using \hypnamref[alg:fgemm]{gemm}}
   \STATE $(P_2,L_2,E_2) = \texttt{PLE}(A_4)$; \COMMENT{recursively}
\end{minipage}
\begin{minipage}{0.36\columnwidth}
Split $A$ columnwise in halves: 
$A=\begin{bmatrix}A_1&A_2\end{bmatrix}$\\
Split $A_2=\begin{bmatrix}A_{3}\\A_4\end{bmatrix}$,
$L_1=\begin{bmatrix}L_{1,1}\\L_{1,2}\end{bmatrix}$ 
where $A_3$ and $L_{1,1}$ have $r_1$ rows.
\end{minipage}
\end{minipage}
   \RETURN ($
   P_1   \begin{bmatrix}   I_{r_1}\\&P_2  \end{bmatrix}, 
   \begin{bmatrix}
   L_{1,1}\\P_2L_{1,2}&L_2
   \end{bmatrix}, 
   \begin{bmatrix}
   E_1&A_3\\&E_2
   \end{bmatrix}
$);
\ENDIF
\end{algorithmic}

\end{algorithm}

\subsubsection{Echelon forms}

The row-echelon and reduced row-echelon forms can be obtained from the PLE
decomposition, using additional operations: \texttt{trsm}, \texttt{trtri}
and \texttt{trtrm}, as shown in algorithm~\ref{alg:rowech} and~\ref{alg:redrowech}.

\begin{algorithm}{[\texttt{RowEchelon}\label{alg:rowech}]}
\begin{algorithmic}[1]
\label{alg:echelon}
\REQUIRE{ $A \in \F_q^{m \times n}$ }
\ENSURE{$(X,E)$ such that $XA=E$, $X$ is non-singular and  $E$ is in row-echelon
form}
\STATE $(P,L,E) = \texttt{PLE}(A)$; \\
\begin{minipage}{\columnwidth}
\begin{minipage}{0.58\columnwidth}
\STATE $X_1=L_1^{-1}$; \COMMENT{using \hypnamref[alg:trtri]{trtri}}
\STATE $X_2=-L_2X_1$; \COMMENT{using \hypnamref[alg:trmm]{trmm}}
\end{minipage}\hfill
\begin{minipage}{0.36\columnwidth}
Split $L = 
\begin{bmatrix}
L_1\\L_2
\end{bmatrix}$, $L_1: r\times r$.
\end{minipage}
\end{minipage}
\RETURN  $\left(X = 
\begin{bmatrix}
X_1\\X_2&I_{m-r}\end{bmatrix} P^T, E\right);$
\end{algorithmic}
\end{algorithm}

\begin{algorithm}{[\texttt{ReducedRowEchelon}\label{alg:redrowech}]}
\begin{algorithmic}[1]
\label{alg:redechelon}
\REQUIRE{ $A \in \F_q^{m \times n}$ }
\ENSURE{$(Y,R)$ such that $YA=R$, $Y$ is non-singular and  $R$ is in reduced row-echelon
form}
\STATE $(X,E) = \texttt{RowEchelon}(A)$;
\STATE Let $Q$ be the permutation matrix that brings the leading row
coefficients of E to the diagonal;
\STATE Set $EQ=
\begin{bmatrix}
U_1&U_2\end{bmatrix}$; \COMMENT{where $U_1$ is  $r\times r$ upper triangular}
\STATE $Y_1=U_1^{-1}$; \COMMENT{using \hypnamref[alg:trtri]{trtri}}
\STATE $Y_1=Y_1X_1$; \COMMENT{using  \hypnamref[alg:trtrm]{trtrm}}
\STATE $R = 
\begin{bmatrix}
I_r&U_1^{-1}U_2
\end{bmatrix} Q^T$; \COMMENT{using \hypnamref[alg:trsmrec]{trsm}}
\RETURN $\left(Y=
\begin{bmatrix}
Y_1\\U_2&I_{n-r}
\end{bmatrix} P^T, R\right) ;$
\end{algorithmic}
\end{algorithm}

\begin{figure}
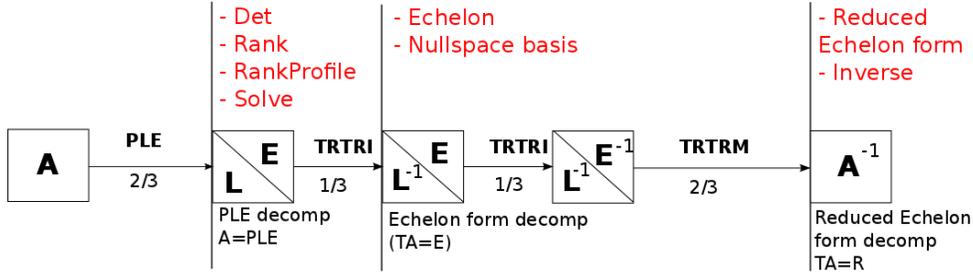

\begin{center}
\includereduc
\end{center}
\caption{Reductions from PLE decomposition to Reduced echelon form}\label{fig:reductions}
\end{figure}

Figure~\ref{fig:reductions} shows the various steps between the classical
Gaussian elimination (LU decomposition), the computation of the echelon form and
of the reduced echelon form, together with the various problems that each of
them solve. Table~\ref{tab:gaussconstants} shows the leading constant $K_\omega$ in the
asymptotic time complexity of these algorithms, assuming that two $n\times n $
matrices can be multiplied in $C_\omega n^\omega + o(n^\omega)$.
\begin{table}
    \begin{tabular}{llll}
%   \toprule
    Algorithm  & Constant $K_\omega$ & $K_3$ & $K_{\log_27}$\\
        \hline
    \texttt{gemm}  & $C_\omega$& 2 & 6 \\
    \texttt{trsm} & $\frac{C_\omega}{2^{\omega-1}-2}$ & $1$  & $4$\\
    \texttt{trtri}& 
    $\frac{C_\omega}{(2^{\omega-1}-2)(2^{\omega-1}-1)}$&$\frac{1}{3}\approx 0.33$&$\frac{8}{5}=1.6$\\
      \texttt{trtrm}, \texttt{PLE}
    &
    $\frac{C_\omega}{2^{\omega-1}-2}-\frac{C_\omega}{2^{\omega}-2}$ &
    $\frac{2}{3}\approx 0.66$ & $\frac{14}{5} =2.8$ \\
    \texttt{Echelon}&  $\frac{C_\omega}{2^{\omega-2}-1}-\frac{3C_\omega}{2^{\omega}-2}$ & 1 &
      $\frac{22}{5}\approx 4.4$\\
    \texttt{RedEchelon} &$\frac{C_\omega(2^{\omega-1}+2)}{(2^{\omega-1}-2)(2^{\omega-1}-1)}$& 2 &
    $\frac{44}{5}= 8.8$\\
  %\texttt{GJ$^*$} &$\frac{C_\omega}{2^{\omega-2}-1}$&2&8&$\times$\\
 %   \hline
\end{tabular}
\caption{Complexity of elimination algorithms}\label{tab:gaussconstants}
\end{table}

\begin{remark}
Note that, if the rank $r$ is very small compared to the dimensions $m\times n$ of
the matrix, a system $A x = b$ can be solved in time bounded by
$\BigO{(m+n)r^2}$ \cite[Theorem~1]{MR1805128}. 
\end{remark}

\subsection{Minimal and characteristic polynomial of a dense matrix}
\index{Characteristic polynomial}
\index{Minimal polynomial}

%%%%%%%%%%%%%%%%%%
\begin{definition}
\begin{enumerate}
\item A {\em Las-Vegas algorithm} is a randomized algorithm which is always
  correct. Its expected running is time is always finite.
\item A {\em Monte-Carlo algorithm} is a randomized algorithm which is correct with a
   certain probability. Its running time is deterministic.
\end{enumerate}
\end{definition}

The computation of the minimal and characteristic polynomials is closely related
to that of the Frobenius normal form.
%%%%%%%%%%%%%%%%%%
\begin{definition} Any matrix $A \in \F_q^{n\times n}$ is similar to a
  unique block diagonal matrix $F=P^{-1} A P =
  diag(C_{f_1},\ldots,C_{f_t})$ where the blocks $C_{f_i}$ are
  companion matrices of the polynomials $f_i$, which satisfy
  $f_{i+1}|f_i$. The $f_i$ are the {\em invariant factors} of
  $A$ and $F$ is the {\em Frobenius normal form} of $A$. 
\end{definition}
Most algorithms computing the minimal and characteristic polynomial or the
Frobenius normal form rely on Krylov basis computations.
%%%%%%%%%%%%%%%%%%
\begin{definition}
\begin{enumerate}
\item The Krylov matrix of order $d$ for a vector $v$ w.r.t a matrix $A$ is the
  matrix 
$K_{A,v,d}=
\begin{bmatrix}
 v&Av&\dots&A^{d-1}v
 \end{bmatrix} \in \F_q^{n\times d}$.
\item The minimal polynomial $P_\text{min}^{A,v}$ of $A$ and $v$
 is the least degree monic polynomial $P$ such that $P(A)v=0$.
\end{enumerate}
\end{definition}

%%%%%%%%%%%%%%%%%%
\begin{theorem}
\begin{enumerate}
\item $AK_{A,v,d}=K_{A,v,d}C_{P_\text{min}^{A,v}}$, where
  $d=\deg(P_\text{min}^{A,v})$.
\item For lineraly independent vectors $(v_1,\dots,v_k)$, if   $K=\begin{bmatrix}K_{A,v_1,d_1}&\dots& K_{A,v_k,d_k}\end{bmatrix}$ is non
  singular. Then $AK=
K\begin{bmatrix}
C_{P_\text{min}^{A,v_1}}&B_{1,2}&\dots&B_{1,k}\\
B_{2,1}&C_{P_\text{min}^{A,v_1}}&\dots&B_{2,k}\\
\vdots&\vdots&\ddots&\vdots\\
B_{k,1}&B_{k,2}&&C_{P_\text{min}^{A,v_k}}
\end{bmatrix}$, where the blocks $B_{i,j}$ are zero except on the last column.
\item For linearly independent vectors $(v_1,\dots,v_k)$, let $(d_1,\dots
  d_k)$ be the lexicographically largest sequence of degrees such that
  $K=\begin{bmatrix}K_{A,v_1,d_1}&\dots& K_{A,v_k,d_k}\end{bmatrix}$ is non-singular. Then 
\begin{equation}
K^{-1}AK=
\begin{bmatrix}
C_{P_\text{min}^{A,v_1}}&B_{1,2}&\dots&B_{1,k}\\
&C_{P_\text{min}^{A,v_1}}&\dots&B_{2,k}\\
&&\ddots&\vdots\\
&&&C_{P_\text{min}^{A,v_k}}
\end{bmatrix}=H
\label{eq:hessenberg}
\end{equation}
\end{enumerate}
\end{theorem}

%%%%%%%%%%%%%%%%%%
\begin{remark}
\begin{enumerate}
\item Some choice of vectors $v_1,\dots, v_k$ lead to a matrix $H$ block
  diagonal: this is the Frobenius normal form~\cite{MR1657129}
\item The matrix obtained at equation (\ref{eq:hessenberg}) is called a Hessenberg
  form. It suffices to compute  the characteristic polynomial from its diagonal blocks.
\end{enumerate}
\end{remark}

%%%%%%%%%%%%%%%%%%%
\begin{theorem}
 The Frobenius normal form can be computed:
\begin{enumerate}
\item\label{alg:charp:det} by a deterministic algorithm~\cite{StoVil00} in $6n^3+\BigO{n^2\log^2n}$ field
  operations,(only $(2+\frac{2}{3})n^3+\BigO{n^2}$ for the characteristic polynomial~\cite{MR2280540})
\item by a deterministic algorithm~\cite{MR1948725} in $\BigO{n^\omega\log n \log\log
  n}$, together with a transformation matrix, (only $\BigO{n^\omega\log n}$ for the characteristic polynomial~\cite{MR796306} )
\item by a Las-Vegas algorithm~\cite{MR1834824} in $\BigO{n^\omega\log n}$
  field operations for any field, together with a transformation matrix
\item \label{alg:charp:arith} by a Las-Vegas algorithm~\cite{MR2402276} in $\BigO{n^\omega}$ for
  $q>2n^2$, without transformation matrix.
\end{enumerate} 

 The minimal\index{Minimal polynomial} and characteristic\index{Characteristic polynomial} polynomials,  obtained
 as the first invariant factor and the product of all invariant factors, can be
 computed with the same complexities.
\end{theorem}

%%%%%%%%%%%%%%%%%%
\begin{remark}
These algorithms are all based Krylov bases. Algorithm~(\ref{alg:charp:det}.)
iteratively compute the Krylov iterates one after the other. Their cubic time
complexity with a small leading constant makes them comparable to Gaussian
elimination.
A fast exponentiation scheme by Keller-Gehrig~\cite{MR796306} achieves a
sub-cubic time complexity for the characteristic polynomial, off by a logarithmic factor of n from the matrix
multiplication. 
The choice for the appropriate vectors that will generate the Frobenius normal
form can be done either probabilistically (Las-Vegas) or deterministically with an
$\log\log n$ factor.
% In order to remove the $\log n$ factor and match the lower bound complexity of
% matrix multiplication, 
Algorithm~(\ref{alg:charp:arith}.) uses a
different iteration where the size of the Krylov increases according to an
arithmetic progression rather than geometric (as all others) and the transformation matrix is not
computed. This allows it to match to the complexity of matrix
multiplication. This reduction is practical and is implemented as in {\sc
  LinBox}.
\end{remark}

\begin{remark}\label{rem:extensions}
These probabilistic algorithms depend on the ability to sample uniformly from a
large set of coefficients from the field.
Over small fields, it is always possible to embed the problem into an
extension field, in order to make the random sampling set
sufficiently large. In the worst case, this could add a
$\BigO{\log(n)}$ factor to the arithmetic cost and prevent most of the
bit-packing techniques.
Instead, the effort of~\cite{MR1834824} is to handle cleanly the small finite
field case.
\end{remark}

\subsection{Blackbox iterative methods}\label{ssec:blackbox}
We consider now the case where the input matrix is sparse, i.e., has
many zero elements, or has a structure which enables fast
matrix-vector products. Gaussian elimination would fill-in the sparse
matrix or modify the interesting structure. Therefore one can use
iterative methods instead which only use matrix-vector iterations
({\em blackbox methods} \cite{MR1056629}). 
There are two major differences with numerical iterative 
routines: over finite fields there exists isotropic vectors and there is no
notion of convergence, hence the iteration must proceed until exactness of the result~\cite{Lam96}.
Probabilistic early termination can nonetheless be applied when the
degree of the minimal polynomial is smaller than the dimension of the
matrix~\cite{MR1687279,DumVil02,Ebe03}. More generally the probabilistic nature
of the algorithms presented in this section is subtle: e.g., the computation of
the minimal polynomial is Monte-Carlo, but that of system solving, using the
minimal polynomial, is Las-Vegas (by checking consistency of the produced
solution with the system). Making some of the Monte-Carlo solutions Las-Vegas is
a key open-problem in this area.
 
\subsubsection{Minimal\index{Minimal polynomial} Polynomial and the Wiedemann algorithm
 }\index{Wiedemann}
The first iterative algorithm and its analysis are due to D. Wiedemann
\cite{MR831560}. The algorithm computes the minimal\index{Minimal polynomial}
polynomial in the Monte-Carlo probabilistic fashion.
\begin{definition} For a linearly recurring sequence $S=(S_i)$, its
  minimal\index{Minimal polynomial} polynomial is denoted by $\Pi_S$.
\begin{itemize}
\item The minimal\index{Minimal polynomial} polynomial of a matrix is denoted $\Pi_A = \Pi_{(A^i)}$.
\item For a matrix $A$ and a vector $b$, we note
  $\Pi_{A,b}=\Pi_{(A^i\cdot b)}$.
\item With another vector $u$, we note
  $\Pi_{u,A,b}=\Pi_{(u^T\cdot A^i \cdot b)}$.
\end{itemize}
\end{definition}

\begin{algorithm}{[Wiedemann minimal\index{Minimal polynomial} polynomial]\label{alg:wiedemann}}
\begin{algorithmic}[1]
\REQUIRE $A \in \F_q^{n\times n}$, $u, b \in \F_q^n$.
\ENSURE $\Pi_{u,A,b}$.
\STATE Compute $S=(u^T A^i b)$ for $i \leq 2n$;
\STATE Use the Berlekamp-Massey algorithm to compute the minimal\index{Minimal polynomial}
polynomial of the scalar sequence $S$;
\end{algorithmic}
\end{algorithm}

\begin{definition}\seedefTotient
We extend Euler's totient function by 
$\Phi_{q,k}(f)=\prod (1-q^{-kd_i}),$ where ${d_i}$ are the degrees of the distinct monic
irreducible factors of the polynomial~$f$.
\index{function!Euler's $\Phi$}
\end{definition}

\begin{theorem} For vectors $u_1, \ldots, u_j$ selected uniformly at random,
  the probability that $\operatorname{lcm}(\Pi_{u_j,A,b})=\Pi_{A,b}$ is at
  least $\Phi_{q,k}(\Pi_{A,b})$.
\end{theorem}

\begin{theorem} For vectors $b_1, \ldots, b_k$ selected
  uniformly at random, the probability that
  $\operatorname{lcm}(\Pi_{A,b_i})=\Pi_A$ is at least
  $\Phi_{q,k}(\Pi_A)$.
\end{theorem}
\subsubsection{Rank\index{Rank}, Determinant and
  Characteristic\index{Characteristic polynomials} Polynomial}
It is possible to compute the rank\index{Rank}, determinant, and
characteristic\index{Characteristic polynomials} polynomial of a
matrix from its minimal\index{Minimal polynomial} polynomial. All
these reductions require to precondition the matrix so that the
minimal\index{Minimal polynomial} polynomial of the obtained matrix
will reveal the information sought, while keeping a low cost for the
matrix-vector product
\cite{MR1809985,MR1229306,DumVil02,Tur02,MR1849765,Vil03,MR1878939}. 
\begin{theorem}\cite{MR1809985}\label{thm:EK} 
  Let S be a finite subset of a field $\F$ that does not include~$0$. 
  Let $A \in \F^{m \times n}$ having rank\index{Rank} $r$. 
  Let $D_1 \in S^{n \times n}$ and $D_2 \in S^{m \times m}$ be two
  random diagonal matrices then 
  $degree(minpoly( D_1 \times A^t \times D_2 \times A \times D_1 ))=r$, 
  with probability at least $1 - \frac{11.n^2-n}{2|S|}$.
\end{theorem}

\begin{theorem}\cite{Tur02}\label{thm:det} 
  Let S be a finite subset of a field $\F$ that does not include~$0$. 
  Let $U\in S^{n \times n}$ be a unit upper bi-diagonal matrix where
  the second diagonal elements $u_1,\ldots,u_{n-1}$ are randomly
  selected in $S$.
  For $A \in \F^{n \times n}$, the term of degree $0$ of the
  minimal\index{Minimal polynomial} polynomial of $UA$ is the
  determinant of $A$ with probability at least $1-\frac{n^2-n}{2|S|}$.
\end{theorem}
\begin{remark}
If $A$ is known to be non-singular the algorithm can be repeated with
different matrices $U$ until the obtained minimal polynomial is of
degree $n$. Then it is the characteristic polynomial of $UA$ and the
determinant is certified. 
Alternatively if the matrix is singular then $X$ divides the minimal
polynomial. As Wiedemann's algorithm always returns a factor of the
true minimal polynomial, and $U$  is invertible,
the algorithm can be repeated on $UA$ until either the obtained
polynomial is of degree $n$ or it is divisible by $X$.
Overall
the determinant has a Las-Vegas blackbox solution.
\end{remark}

\begin{theorem}\cite{MR1849765,Vil03}\label{thm:rankupdate}
  Let S be a finite subset of a field $\F$
  that does not include~$0$ and $A\in \F^{n\times n}$ with $s_1,
  \ldots, s_t$ as invariant factors. Let $U \in S^{n\times k}$ and $V
  \in S^{k\times n}$ be randomly chosen rank\index{Rank} $k$ matrices
  in $\F$. Then $\operatorname{gcd}(\Pi_A, \Pi_{A+UV})=s_{k+1}$ with
  probability at least $1 -\frac{nk + n + 1}{|S|}$.
\end{theorem}
\begin{remark}
Using the divisibility of the invariant factors and the fact that
their product is of degree $n$, one can see that the number of degree
changes between successive invariant factors is of order
$\BigO{\sqrt{n}}$~\cite{MR1849765}. Thus by a binary search over
successive applications of theorem \ref{thm:rankupdate} one can
recover all of the invariant factors and thus the
characteristic\index{Characteristic polynomials} polynomial of the
matrix in a Monte-Carlo fashion. 
\end{remark}
\subsubsection{System solving and the Lanczos algorithm}\index{Lanczos}
For the solution of a linear system $Ax=b$, one could compute the
minimal\index{Minimal polynomial} polynomial $\Pi_{A,b}$ and then derive a solution of the
system as a linear combination of the $A^ib$. 
The following Lanczos approach is more efficient for system solving as
it avoids recomputing (or storing) the latter vectors
\cite{MR1809985,MR1805174}.
\begin{algorithm}{[Lanczos system solving]\label{alg:lanczos}}
\begin{algorithmic}[1]
\REQUIRE $A \in \F^{m\times n}$, $b \in \F^m$.
\ENSURE $x\in \F^n$ such that $Ax=b$ or {\em failure}.
\STATE\label{lin:datdad} Let $\tilde{A}=D_1 A^T D_2 A D_1$ and $\tilde{b} = D_1 A^T D_2
b+\tilde{A} v$ with $D_1$ and $D_2$ random diagonal matrices and $v$ a
random vector;
\STATE $w_0=\tilde{b}$; $v_1=\tilde{A}w_0$; $t_0=v_1^T w_0$;
$\gamma=\tilde{b}^tw_0 t_0^{-1}$; $x_0 = \gamma w_0$;
\REPEAT
\STATE $\alpha=v^T_{i+1}v_{i+1} t_i^{-1}$; $\beta=v^T_{i+1}v_{i}
t_{i-1}^{-1}$; $w_{i+1} = v_{i+1} -\alpha w_i -\beta w_{i-1}$;
\STATE $v_{i+2}= \tilde{A} w_{i+1}$; $t_{i+1}=w_{i+1}^T v_{i+2}$;
\STATE $\gamma=\tilde{b}^tw_{i+1}t_{i+1}^{-1}$; $x_{i+1}=x_i + \gamma w_{i+1}$;
\UNTIL{$w_{i+1}=0$ or $t_{i+1}=0$;}
\STATE Return $x=D_1(x_{i+1}-v)$;
\end{algorithmic}
\end{algorithm}

The probability of success of algorithm \ref{alg:lanczos} follows also
theorem \ref{thm:EK}. 
\begin{remark}
Over small fields, if the rank of the matrix is known, the
diagonal matrices of line \ref{lin:datdad} can be replaced by sparse
preconditioners with $\BigO{n\log(n)}$ non-zero coefficients to avoid the
need of field extensions\cite[corollary 7.3]{MR1878939}. 
\end{remark}
\begin{remark}
If the system with $A$ and $b$ is known to have a solution then the
algorithm can be turned Las-Vegas by checking that the output $x$
indeed satisfies $Ax=b$. In general, we do not know if this algorithm
returns failure because of bad random choices or because the system is
inconsistent. However, Giesbrecht, Lobo and Saunders have shown that
when the system is inconsistent, it is possible to produce a
certificate vector $u$ such that $u^T A=0$ together with 
$u^T b \neq 0$ within the same complexity 
\cite[Theorem 2.4]{MR1805174}. Overall, system solving can be
performed by blackbox algorithms in a Las-Vegas fashion.
\end{remark}
\subsection{Sparse and structured methods}\index{Sparse matrix}
\index{Structured matrix}\index{Gaussian elimination}\index{Fill-in}
Another approach to sparse linear system is to use Gaussian
elimination with pivoting, taking into account the zero coefficients. 
This algorithm modifies the structure of the matrix and might suffer
from fill-in. Consequently the available memory is usually the
bottleneck. From a triangularization one can naturally derive the
rank\index{Rank}, determinant, system solving and nullspace. 
Comparisons with the blackbox approaches above can be found e.g., in
\cite{DumVil02}.
\subsubsection{Reordering}\index{Reordering}
\begin{algorithm}{[Gaussian elimination with linear pivoting]\label{alg:reord}}
\begin{algorithmic}[1]
\REQUIRE a matrix $A \in \F^{m \times n}$;
\ENSURE An upper triangular matrix $U$ such that there exists a
unitary lower-triangular matrix $L$ and permutations matrices $P$ and
$Q$ over $\F$, with $A=P\cdot L \cdot U \cdot Q$;
\FORALL{elimination steps}
\STATE Choose as pivot row the sparsest remaining row;
\STATE In this row choose the non zero pivot with lowest number of non
zero elements in its column;
\STATE Eliminate using this pivot;
\ENDFOR
\end{algorithmic}
\end{algorithm}

\begin{remark} 
Yannakakis showed that finding the minimal fill-in (or
equivalently the best pivots) during
Gaussian elimination is an NP-complete task
\cite{MR604513}. 
In numerical algorithms, heuristics have been developed and  
comprise minimal degree ordering, cost functions or nested dissection
(see
e.g.,
\cite{MR1135327,MR2124398,MR1642639}).
These heuristics for reducing fill-in in the numerical setting, often
assume symmetric and invertible matrices, and do not take into account
that new zeros may be produced by elimination operations ($a_{ij} =
a_{ij} + \delta_i * a_{kj}$), as is the case with matrices over finite
fields. \cite{DumVil02} thus proposed the heuristic \ref{alg:reord} to
take those new zeros into account, using a local optimization of a
cost function at each elimination step.
\end{remark}

\subsubsection{Structured matrices and displacement
  rank\index{Displacement rank}}
Originating from the seminal paper \cite{MR537629} most of the
algorithms dealing with structured matrices use the displacement rank\index{Displacement rank}
approach \cite{MR1843842}.
\begin{definition}
For $A\in \F^{m\times m}$ and $B\in\F^{n\times n}$, the {\em Sylvester
  (resp. Stein)
linear displacement operator $\bigtriangledown_{A,B}$
(resp. $\bigtriangleup_{A,B}$)} satisfy for $M\in\F^{m\times n}$:
\begin{align*}
\bigtriangledown_{A,B}(M) = AM - MB\\
\bigtriangleup_{A,B}(M) = M - AMB
\end{align*}
A pair of matrices $(Y,Z)\in\F^{m\times \alpha} \times \F^{n\times
  \alpha}$ is a {\em $A,B$-Sylvester-generator\index{Sylvester
    generator} of length $\alpha$} (resp. Stein\index{Stein
  generator}) for $M$ if $\bigtriangledown_{A,B}(M) = Y Z^T$
(resp. $\bigtriangleup_{A,B}(M) = Y Z^T$). 
\end{definition}
The main idea behind algorithms for structured matrices is to use such
generators as a compact data structure, in cases where the
displacement has low rank\index{Rank}. 

Usual choices of matrices $A$ and $B$ are diagonal matrices and
cyclic down shift matrices:
\begin{definition}
$\mathbb{D}_x, x\in\F^n$ is the diagonal matrix whose $(i,i)$
  entry is $x_i$.\\
$\mathbb{Z}_{n,\varphi},\varphi\in\F$ is the $n\times n$ unit
  circulant matrix having $\varphi$ at position $(1,n)$, ones in the
  subdiagonal $(i+1,i)$ and zeros elsewhere.
\end{definition}

\begin{table}[htbp]
\begin{tabular}{|c|c|c|c|c|}
%\hline
\multicolumn{2}{|c|}{operator matrices}& class of structured & rank\index{Rank} of
& number of flops \\
A & B & matrices $M$ & $\bigtriangledown_{A,B}(M)$ & for computing $M
\cdot v$\\
\hline
$\mathbb{Z}_{n,1}$& $\mathbb{Z}_{n,0}$& Toeplitz\index{Toeplitz matrix} and its inverse& $\leq
2$& $\BigO{(m+n)\log(m+n)}$\\
$\mathbb{Z}_{n,1}$& $\mathbb{Z}_{n,0}^T$& Hankel\index{Hankel matrix} and its inverse& $\leq 2$& $\BigO{(m+n)\log(m+n)}$\\
$\mathbb{Z}_{n,0}+\mathbb{Z}_{n,0}^T$&$\mathbb{Z}_{n,0}+\mathbb{Z}_{n,0}^T$ & Toeplitz\index{Toeplitz matrix}+Hankel\index{Hankel matrix}& $\leq 4$& $\BigO{(m+n)\log(m+n)}$\\
$\mathbb{D}_{x}$& $\mathbb{Z}_{n,0}$& Vandermonde\index{Vandermonde matrix}& $\leq 1$& $\BigO{(m+n)\log^2(m+n)}$\\
$\mathbb{Z}_{n,0}$& $\mathbb{D}_{x}$ & inverse of Vandermonde\index{Vandermonde matrix}& $\leq 1$& $\BigO{(m+n)\log^2(m+n)}$\\
$\mathbb{Z}_{n,0}^T$& $\mathbb{D}_{x}$ & transposed of Vandermonde\index{Vandermonde matrix}& $\leq 1$& $\BigO{(m+n)\log^2(m+n)}$\\
$\mathbb{D}_{x}$& $\mathbb{D}_{y}$ & Cauchy\index{Cauchy matrix} and its inverse& $\leq 1$& $\BigO{(m+n)\log^2(m+n)}$\\
%\hline
\end{tabular}
\caption{Complexity of the matrix-vector product for some structured matrices}\label{tab:struct}
\end{table}
As computing matrix vector products with such structured matrices have
close algorithmic correlation to computations with polynomials and
rational functions, these matrices can be multiplied by vectors fast,
in nearly linear time as shown on table \ref{tab:struct}. 
Therefore the algorithms of section \ref{ssec:blackbox} can naturally
be applied to structured matrices, to yield almost $\BigO{n^2}$ time
linear algebra.

Now, if the displacement rank\index{Displacement rank} is small there exists 
algorithms quasi linear in $n$, the dimension of the matrices, which
over finite fields are essentially variations or extensions of the
Morf/Bitmead-Anderson divide-and-conquer \cite{Mor80,MR591427} or
Cardinal's \cite{MR1696212} approaches.
The method is based on dividing the original problem repeatedly into
two subproblems with one leading principal submatrix and the related
Schur complement. This leads to $\BigO{\alpha^2n^{1+o(1)}}$ system solvers,
which complexity bound have recently been reduced to
$\BigO{\alpha^{\omega-1}n^{1+o(1)}}$ \cite{MR2463004,JeaMou10}.
We few exceptions, all algorithms thus need matrices in generic rank
profile. Over finite fields this can be achieved using Kaltofen and
Saunders unit upper triangular Toeplitz preconditioners
\cite{MR1229306}  and by controlling the displacement rank growth and
non-singularity issues \cite{Kal94}. 

\subsection{Hybrid methods}
\subsubsection{Hybrid sparse-dense methods}
Overall, as long as the matrix fits into memory, Gaussian elimination
methods are usually faster than iterative methods, over finite fields
\cite{DumVil02}.
There are then heuristics trying to take the best of both
strategies. Among those we briefly mention the most widely used:
\begin{itemize}
\item Perform the Gaussian elimination with reordering
  \ref{alg:reord} until the matrix is almost filled-up. If the
  remaining non-eliminated part would fit as a dense matrix, switch to
  the dense methods of section \ref{ssec:echelon}.
\item Maintain two sets of rows (or columns), sparse and
  dense. Favor elimination on the sparse set. This is
  particularly adapted to index calculus \cite{LamOdl91}. 
\item Perform a preliminary reordering in order to cut the matrix into
  four quadrants, the upper left one being triangular. This, together
  with the above strategies has proven effective on matrices which are
  already quasi-triangular, e.g., Gr{\"o}bner bases
  computations in finite fields \cite{FauLac10}.
\item If the rank\index{Rank} is very small compared to the dimension of the
  matrix, one can use left and right highly rectangular projections to
  manipulate smaller structures \cite{MR2402272}.
\item The arithmetic cost and thus timing predictions are easier on
  iterative methods than on elimination methods. On the other hand the
  number of non-zero elements at a given point of the elimination is
  usually increasing during an elimination, thus providing a lower
  bound on the remaining time to triangularize. Thus a heuristic is to
  perform one matrix-vector product with the original matrix and then
  eliminate using Gaussian elimination. If at one point the lower
  bound for elimination time surpasses to predicted iterative one or
  if the the algorithm runs out of memory, stop the elimination and
  switch to the iterative methods \cite{DurSauWan03}.
\end{itemize}

\subsubsection{Block-iterative methods}

Iterative methods based on one-dimensional projections, such as Wiedmann and
Lanczos algorithm can be generalized with block projections.
Via efficient preconditioning \cite{MR1878939} these extensions to the
scalar iterative methods can present enhanced properties:
\begin{itemize}
\item Usage of dense sub-blocks, after multiplications of blocks of
  vectors with the sparse matrix or the blackboxes, allows for a
  better locality and optimization of memory accesses, via the
  application of the methods of section~\ref{ssec:blas}.
\item Applying the matrix to several vectors simultaneously introduces more
  parallelism \cite{MR1236735,MR1192970,MR1687279}.
\item Also, their probability of success augments with the size of the
  considered blocks, especially over small fields \cite{MR1270621,Vil97}.
\end{itemize}

\begin{definition}
Let $X\in \F_q^{k\times n}$, $Y\in \F_q^{n\times k}$ and $H_i =XA^iY$ for
$i=0\dots n/k$.
The matrix minimal\index{Minimal polynomial} polynomial of the sequence $H_i$ is the matrix polynomial
$F_{X,A,Y} \in \F_q[X]^{k\times k}$ of least degree, with its leading degree
matrix column-reduced, that annihilates the sequence $(H_i)$.
\end{definition}

\begin{theorem}
The degree $d$ matrix minimal\index{Minimal polynomial} polynomial of a block sequence $(H_i) \in (F_q^{k\times
k})^\Z$ can be computed in $\BigO{k^3d^2}$ using block versions of Hermite-Pade
approximation and extended Euclidean algorithm~\cite{MR1779720} or Berlkamp-Massey
algorithm~\cite{MR1192970,MR1270621,Vil97}. Further improvement
by~\cite{MR1779720,MR2049765,MR2035204,MR2120701}  bring this complexity down
to $\BigO{k^\omega d}^{1+o(1)}$, using a matrix extended Euclidean algorithm.
\end{theorem}

\begin{algorithm}{[Nullspace vector]\label{alg:nullspace}}
\begin{algorithmic}
\REQUIRE{ $A\in \F_q^{n\times n}$}
\ENSURE {$\omega \in \F_q^n$ a vector in the nullspace of $A$}
\STATE Pick $X\in \F_q^{k\times n},Y\in \F_q^{n\times k}$ uniformly at random;
\STATE Compute the sequence $H_i=XA^iY$;
\STATE Compute $F_{X,A,Y}$ the matrix minimal\index{Minimal polynomial} polynomial;
\STATE Let $f=f_rx^r+\dots+f_dx^d$ be a column of $F_{X,A,Y}$;
\STATE Return $\omega=Yf_r +AYf_{r+1}+ \dots +A^{d-r}Yf_{d}$;
\end{algorithmic}
\end{algorithm}

\begin{remark}
These block-Krylov techniques are used to achieve the best known time
complexities for several computations with black-box matrices over a finite
field or the ring of integers: computing the
determinant, the characteristic\index{Characteristic polynomials} polynomial~\cite{MR2120701} and the  solution of a linear system
of equations~\cite{MR2396196}.
\end{remark}

\AllRefCited{MR2124398,MR0269546,MR1779720,MR591427,BooBra09,MR2463004,MR0331751,MR1696212,MR1878939,MR1236735,MR1192970,MR1449760,MR1056627,MR2500374,DumFouSal11,DumGauGieGioHovKalSauTurVil02,MR2035233,MR2738206,MR2280540,DumVil02,DurSauWan03,MR1834824,Ebe03,MR2396196,MR1809985,FauLac10,MR1657129,MR1805174,MR2035204,MR2501869,MR1642639,JeaMou10,MR537629,MR1687279,Kal94,MR1270621,MR1229306,MR1056629,MR2120701,MR796306,LamOdl91,Lam96,MR2402272,Mor80,MR1805128,MR1843842,MR2402276,MR1948725,StoVil00,Sto10,MR2049765,Tur02,Vas11,Vil97,MR1849765,Vil03,WhaPetDon01,MR831560,MR0297115,MR604513,MR1135327}

\subsection{Acknowledgment}
We thank an anonymous referee for numerous helpful suggestions that
considerably improved the paper.

\bibliographystyle{babplain}
\bibliography{addon,sec134}
\end{document}